\documentclass[twocolumn]{aastex631}

\usepackage{booktabs}
\usepackage{amsmath}

\begin{document}

\title{The Fastest Path to Discovering the Second Electromagnetic Counterpart to a Gravitational Wave Event}

\correspondingauthor{Ved~G.~Shah}
\email{vedgs2@illinois.edu}

\author[0009-0009-1590-2318]{Ved~G.~Shah}
\affiliation{Center for Interdisciplinary Exploration and Research in Astrophysics, Northwestern University, Evanston, IL, USA}
\affiliation{Department of Astronomy, University of Illinois Urbana-Champaign, Urbana, IL, USA}

\author[0000-0002-2445-5275]{Ryan~J.~Foley}
\affiliation{Department of Astronomy and Astrophysics, University of California Santa Cruz, Santa Cruz, CA 95064, USA}

\author[0000-0001-6022-0484]{Gautham~Narayan}
\affiliation{Department of Astronomy, University of Illinois Urbana-Champaign, Urbana, IL, USA}
\affiliation{Center for AstroPhysical Surveys, National Center for Supercomputing Applications, Urbana, IL, 61801, USA}
\affiliation{Illinois Center for Advanced Studies of the Universe, University of Illinois Urbana-Champaign, Urbana, IL 61801, USA}
 
\begin{abstract}

The discovery of a second electromagnetic counterpart to a gravitational wave event represents a critical goal in the field of multi-messenger astronomy. In order to determine the optimal strategy for achieving this goal, we perform comprehensive simulations comparing two potential paths forward: continuing the current LIGO-Virgo-KAGRA (LVK) observing run, O4, versus temporarily shutting down the detectors for upgrades before beginning the next observing run, O5. Our simulations incorporate current O4 instrument sensitivities and duty cycles, as well as projected configurations for O5, while accounting for variables such as binary neutron star merger rates, system properties, viewing angles, dust extinction, and kilonova (KN) observables. Our results indicate that a KN discovery would occur $125^{+253}_{-125}$~days (middle 50\% interval) sooner in O5 compared to O4, suggesting that extending O4 would lead to faster discovery if the shutdown period between runs is $>$4~months. Moreover, for 88\% of our simulations, continuing O4 results in earlier KN discovery when compared to the expected two-year shutdown between O4 and O5. Given these findings and the critical importance of avoiding a $>$10 year gap between first and second electromagnetic counterpart discoveries, we suggest LVK consider extending O4 operations for as long as feasible prior to shutting down for critical upgrades.

\end{abstract}


\section{Introduction} 

The LIGO-Virgo and LIGO-Virgo-KAGRA collaborations (LVC and LVK, respectively) have discovered, through observations of gravitational waves (GWs), hundreds of compact-binary mergers \citep{GWTC-1, GWTC-2, GWTC-3, GWTC-2.1}.  However, there have only been two binary neutron star (BNS) mergers detected, GW170817 \citep{GW170817} and GW190425 \citep{GW190425}.  An electromagnetic (EM) counterpart was discovered for GW170817 \citep{2017Sci...358.1556C, 2017ApJ...850L...1L, 2017ApJ...848L..16S, 2017ApJ...848L..24V, LIGO_2017:GW_GRB}, AT~2017gfo (a kilonova; KN), allowing the community to observe the KN across all wavelengths and over several weeks 
\citep{2017PASA...34...69A,
2017Natur.551...64A, 
2017Sci...358.1556C, 
2017ApJ...848L..17C,
2017ApJ...848L..29D,
2017Sci...358.1570D,
2017Sci...358.1565E,
2017SciBu..62.1433H,
2017Sci...358.1559K,
2017ApJ...850L...1L,
2017Natur.551...67P,
2018ApJ...852L..30P,
2017Sci...358.1574S,
2017Natur.551...75S,
2017ApJ...848L..27T,
2017Natur.551...71T,
2017PASJ...69..101U, 
2017ApJ...848L..24V}.  
These observations led to major discoveries related to cosmology \citep{2017Natur.551...85A, 2022Univ....8..289B}, high-energy astrophysics \citep{2017ApJ...848L..13A, 2017ApJ...848L..14G}, gravity \citep{2019CQGra..36n3001B}, nuclear physics \citep{2017ApJ...850L..19M, LATTIMER2019167963}, and compact object formation  \citep{2017ApJ...848L..22B, 2017ApJ...848L..30P}.

No EM counterpart was discovered for GW190425  \citep{Coughlin2019, Hosseinzadeh2019, Lundquist2019, Antier2020, Gompertz2020, Boersma2021, Coulter2024, Paek2024, Smartt2024}.
This is in part because the two-dimensional localization was $\sim$10,183~deg$^{2}$ (90\% confidence), including regions covering the Milky Way plane and near the Sun, making follow-up observations difficult.  More critically, most observations were tuned to discover a counterpart similar to AT~2017gfo.  Because EM observers lacked access to prompt parameter measurements, especially the chirp mass, they did not expect the high mass of GW190425, which suggests a particularly faint and red KN \citep{2020MNRAS.494..190F}, and could not properly adjust their searches to match the expected properties of the KN.  Regardless, {\it no} EM counterpart to any GW event has been discovered since 2017.  As a result, there will be a $>$7~year gap between the discoveries of the first and second GW counterparts.

After the discovery of GW170817, the community expected a relatively high BNS merger -- and EM counterpart -- discovery rate.  Initial measurements of the volumetric BNS merger rates with expected increases in GW detector sensitivity suggested more detections during the third GW observing run (O3); one BNS merger (GW190425) and no EM counterparts were discovered in O3.  This excercise was repeated for the current fourth observing run (O4), resulting in predictions of $3_{-3}^{+6}$ BNS mergers and  $1_{-1}^{+4}$ discoverable KNe, respectively \citep{shah2024predictions}.  Thus far, no BNS mergers nor EM counterparts have been detected in O4.

O4 was originally scheduled to conclude at the end of 2024.  It has since been extended to June 9, 2025.  There is a planned $\ge$2~year downtime between O4 and the next observing run, O5.  If this schedule is kept and no counterparts are detected in the final months of O4, there will be more than a decade gap between GW counterpart discoveries.  Such a gap would have severe consequences for the astrophysical and nuclear physics communities that have built human, physical, and software infrastructure to support this science.

Here, we ask the simple question:  If the goal is to discover a second EM counterpart as quickly as possible, is it better to continue O4 or to shut down, improve the LVK infrastructure, and start O5?  The answer depends on the volumetric rate of BNS mergers, the current O4 parameters, the expected O5 parameters, and the length of downtime between O4 and O5.  In the following sections, we describe a simulation approach to answer this question (Section~\ref{section:method}, Section~\ref{section:set-up}), presenting our results in Section~\ref{section:results}.  We present conclusions and recommendations in Section~\ref{section:conclusions} and \ref{section:recommendations} respectively.

\section{Method}\label{section:method}

\begin{figure*}
    \centering
    \includegraphics[width=\linewidth]{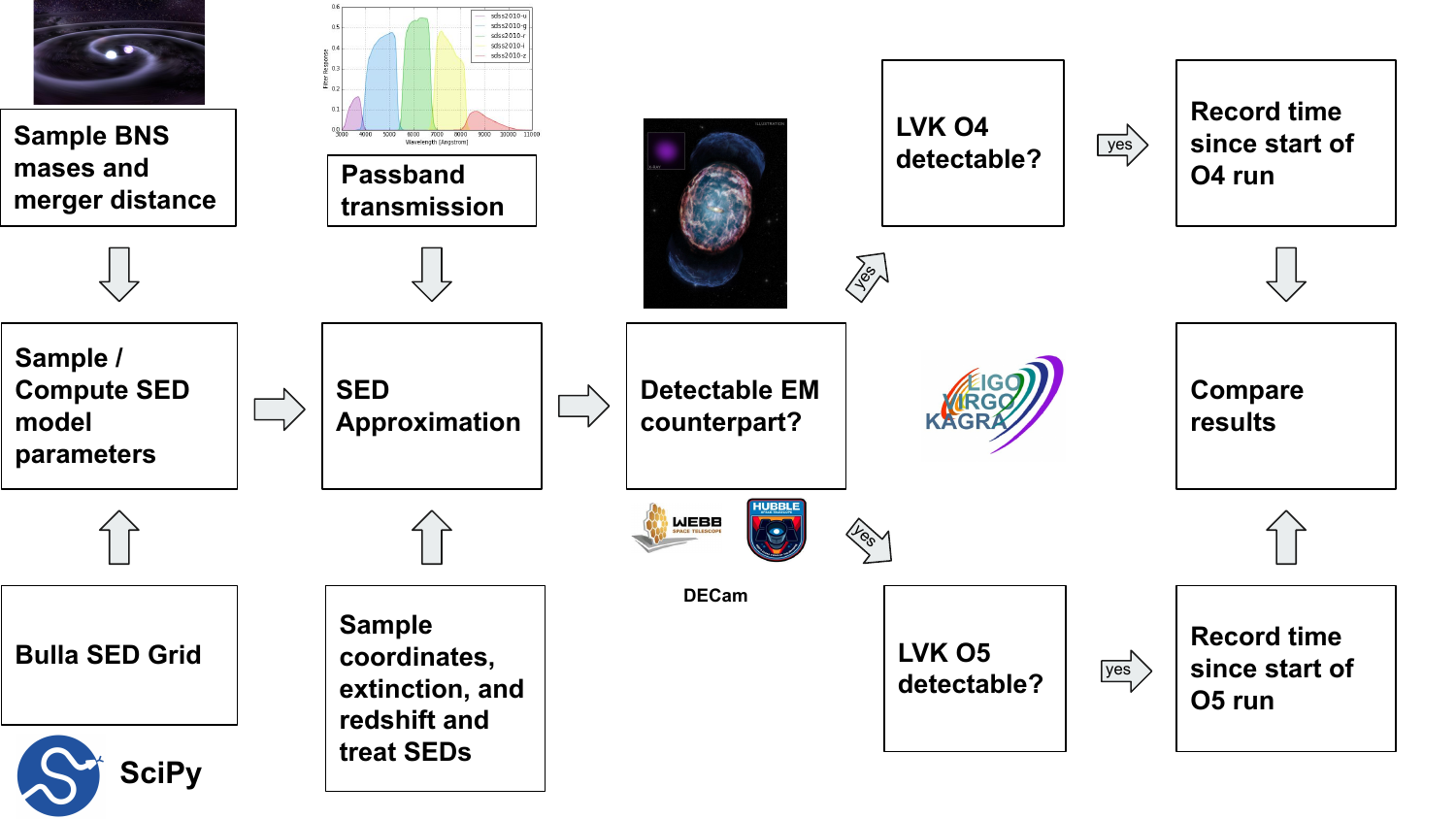}
    \caption{Schematic of the pipeline used to find the time to first KN for LVK O4 and O5. Most components of this pipeline are adapted from the original simulation work done to study discovery rate of KNe \citep{shah2024predictions} (Image Source: NASA, LIGO).}
    \label{figure:new-pipeline}
\end{figure*}

\cite{shah2024predictions} developed a configurable, open-source framework for estimating the rate of discoverable KNe associated with gravitational wave detections of BNS mergers. While we will not recapitulate all the details from the original work here, we summarize the simulation pipeline before detailing changes made to the set up for this work in Section \ref{section:set-up}.

Since running full radiative transfer simulation for each event ($\mathcal{O}(10^6)$ binaries) is computationally infeasible, the authors used interpolation methods over existing KN models \citep{2019MNRAS.489.5037B} to approximate SEDs for each BNS merger. To capture the variance in the population of coalescing NS binaries, the authors sample parameters including, but not limited to, the masses of the merging stars, the host galaxy extinction, the viewing angle, the lanthanide richness of the ejected material, and the distance to the event. A comprehensive list of the input parameters can be found in Table \ref{table:inputs}. These parameters are used, either directly or as inputs to compute secondary parameters, to model both the GW and EM signals from each of the mergers. Since many of the input parameters are sampled from distributions, different combinations of inputs can result in output values that differ significantly. Thus, the authors perform independent Monte Carlo trials until the metrics of interest converge.

While the original code base was developed to answer questions about the number of discoverable KNe over the course of LVK O4 and O5, components of the pipeline are modular and can be re-configured to predict different statistics related to KNe discovery.

We note that while NS-black hole mergers can also produce KNe, these are generally thought to be less common\footnote{\url{https://emfollow.docs.ligo.org/userguide/capabilities.html\#summary-statistics}} \citep{Fragione_2021}. Thus we only consider KNe from BNS mergers.

\section{Experiment Configuration}
\label{section:set-up}

Contained within the aforementioned framework (Section~\ref{section:method}) is a collection of functions that can compute both the GW and EM signals produced by a system of coalescing NSs. In order to facilitate the analysis presented in this work, we build on top of this framework by making some key changes (Figure \ref{figure:new-pipeline}). 

Since we are interested in obtaining an ``apples to apples" comparison for the time to first discoverable KN $(D_{\text{KN}})$ during LVK O4 $(D_{\text{KN}}^{O4})$ and O5 $(D_{\text{KN}}^{O5})$, it is critical to ensure that all other variables, except the ones that are subject to change with the O5 upgrades, remain constant between the O4 and O5 simulations.

We document both the observing run dependent and observing run independent inputs in Table \ref{table:inputs}. Additionally, the values chosen for the detector uptimes and sensitives for LVK O4 and O5 are documented in Tables~\ref{table:duty-cycle} and \ref{table:psd}. We also computed an updated uptime correlation matrix for the LVK instruments using publicly available data from LVK O4b \footnote{\url{https://online.igwn.org/}}. Specifically, we used data from two time periods when LIGO Livingston, LIGO Hanford, and Virgo were operating nominally (April 10, 2024 to July 10, 2024 and August 25, 2024 to October 10, 2024) to calculate the correlation matrix. 

\begin{figure}
    \centering
    \includegraphics[width=\linewidth]{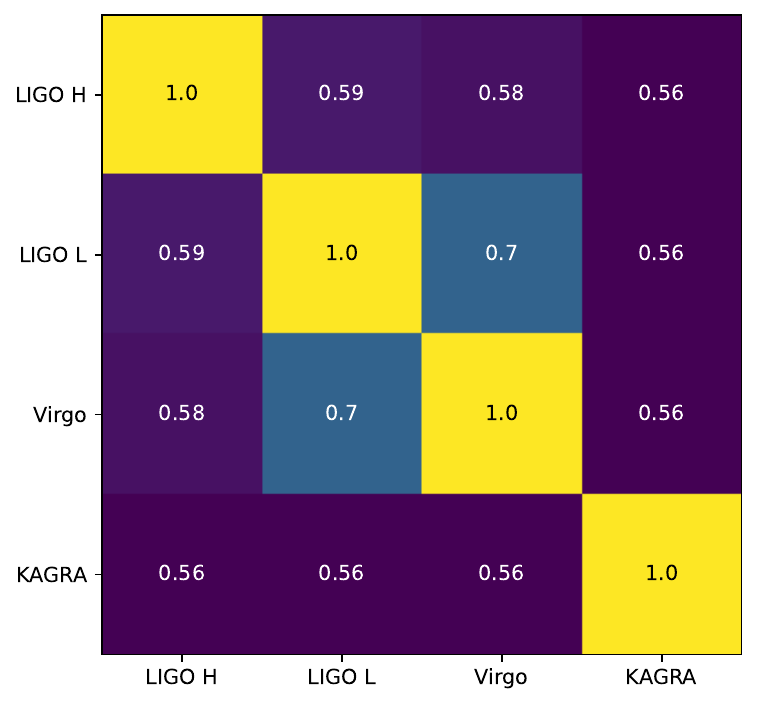}
    \caption{Correlation between the uptimes of different detectors from segments of LVK O3 and O4b  run where the instruments were functioning nominally.}
    \label{figure:correlations}
\end{figure}

\begin{figure}
    \centering
    \includegraphics[width=\linewidth]{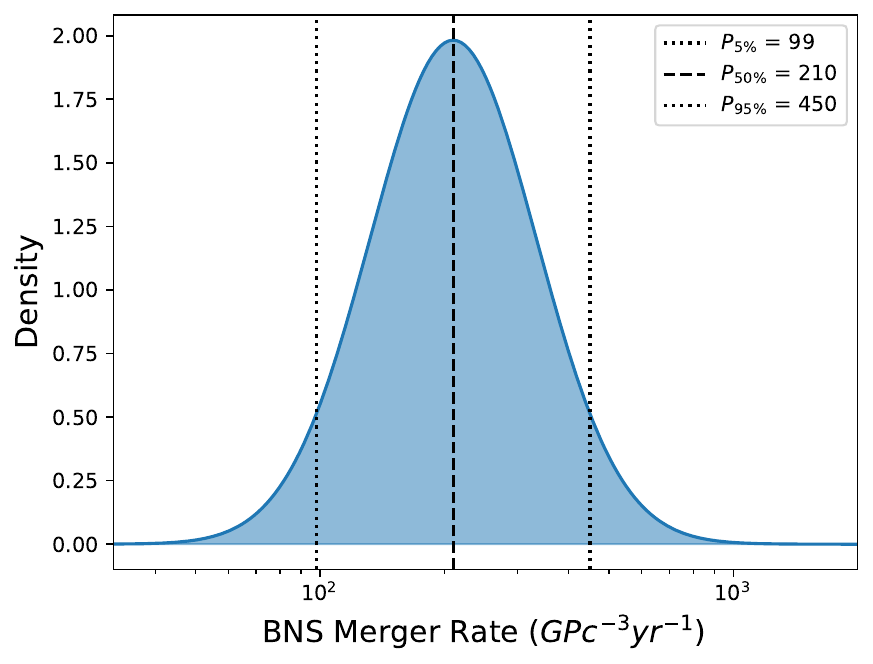}
    \caption{Distribution of BNS merger rates approximates the log normal distribution from the LVK user guide.}
    \label{figure:bns-rates}
\end{figure}

For each trial, we start by sampling from the BNS merger rate distribution (Figure \ref{figure:bns-rates}). Next, we compute the number of events that would take place in a cube of length 910 Mpc since the maximum BNS horizon range for LVK O5 is estimated to be $\sim 450$ Mpc \citep{shah2024predictions}. This volume was specifically chosen to ensure that any BNS mergers outside this simulation volume would not produce a detectable GW for the LVK O5 observing run; i.e., the simulations include all GW events that could possibly be detected. Finally, we compute the number of events over a five-year time period to ensure that there are some BNS mergers in our simulation volume, even when a low BNS merger rate $(\mathcal{R})$ is chosen.  We assume a uniform temporal distribution of the BNS mergers over the 5 year period. Formally, we define the number of events as:

\begin{equation}
    n_{\text{events}} = \mathcal{R} \times (910 \text{ Mpc})^3 \times 5 \text{ yr}
\end{equation}

For each BNS merger, we sample primary parameters and compute the secondary parameters (see Table \ref{table:inputs}) in order to characterize both the gravitational waves and electromagnetic radiation (if any) from the coalescing system. When comparing O4 and O5, with the exception of the run-dependent parameters, the same set of sampled and computed parameters are used for both simulations. Thus, the set of coalescing BNS systems considered for the O4 and O5 configuration for a given trial, are identical. As a result, any difference in the time to first KN detection is exclusively a function of the sensitivity of the LVK instruments and their duty cycles. 

We conducted 1000 independent trials to produce distributions for the number of days to the first KNe discovery for O4. We also produce a distribution for the difference in the time to first detection between LVK O4 and O5, when all run independent variables, including the BNS merger rate, are constant between the two sets of simulations.

\subsection{Discoverability criteria}

For this work, we require that any BNS merger in our simulation is coincidentally detected by at least two LVK instruments with ${\rm SNR} > 8$ and have an EM counterpart with $r_{\rm peak} < 23$~mag (for the DECam $r$ band) to be considered ``discoverable.'' 

We require detections by two GW instruments since a single instrument GW detection for LVK will typically yield very poorly localized skymaps ( $\mathcal{O}(10^4)$~deg$^{2}$) and lower network SNR, both of which make targeted KN searches difficult. The EM criterion is motivated by typical depths reached by the CTIO 4-m telescope with DECam, one of the best instruments for KN discovery, in a reasonable exposure time.

In principle, these criteria are supposed to be the minimum viable criteria for KN detection, and there might be instances when these thresholds are insufficient and others where they are exceeded. Indeed, there may be situations where a KN is discoverable but no detection is made. Reconciling the difference between a discoverable KN and one that is detected requires detailed modeling of the observing capabilities, and inefficiencies \citep{2024arXiv240517558K} of the entire GW follow-up community. Given the single successful detection of a KN with a coincident GW, there currently is not enough data to properly model EM discovery efficiency, and therefore we assume every discoverable KN is detected. 

\begin{table}[]
    \centering
    \begin{tabular}{l|cc}
    \toprule
    \textbf{Instrument} & \textbf{Run O4} & \textbf{Run O5} \\
    \midrule
    LIGO - Livingston & 0.7 & 0.7\\
    LIGO - Hanford & 0.7 & 0.7\\
    Virgo  & 0.47 & 0.7\\
    KAGRA  & 0.27 & 0.7\\
    \bottomrule
    \end{tabular}
    \caption{Duty cycles for detectors used in the LVK O4 and LVK O5 simulations. Figure \ref{figure:correlations} has information on correlation between duty cycles.}
    \label{table:duty-cycle}
\end{table}

\begin{table*}
\centering
\begin{tabular}{lll}
\toprule
\textbf{Instrument} & \textbf{O4 PSD File} & \textbf{O5 PSD File}   \\
\midrule
LIGO - Livingston & aligo\_O4high.txt & AplusDesign.txt \\
LIGO - Hanford  & aligo\_O4high.txt  &  AplusDesign.txt \\ 
Virgo &  avirgo\_O4high\_NEW.txt & avirgo\_O5low\_NEW.txt  \\
KAGRA & KAGRA\_10Mpc.txt  & kagra\_128Mpc.txt   \\

\bottomrule
\end{tabular}
\caption{PSDs for detectors used in the LVK O4 and LVK O5 simulations.}
\label{table:psd}
\end{table*}

\begin{table*}
\centering

\textbf{Observing run independent parameters:}
\bigskip

\begin{tabular}{lll}
\toprule
\textbf{Input(s)} & \textbf{Description}  & \textbf{Reference(s)}\\
\midrule
$x, y, z$ & Cartesian coordinates to the event, with Earth as the center & \cite{shah2024predictions}\\
$m_1$, $m_2$ & Masses of the coalescing NSs & \cite{Galaudage:2020zst}\\
$m_{\text{ej}}^{\text{wind}}$ & Wind ejecta mass from BNS merger & \cite{10.1093/mnras/stad257}\\
$m_{\text{ej}}^{\text{dyn}}$ & Dynamical ejecta mass from BNS merger & \cite{10.1093/mnras/stad257}\\
cos $\Theta$ & Cosine of the observing angle & \cite{2019MNRAS.489.5037B, shah2024predictions}\\
$\phi$ &  Half-opening angle at which the lanthanide-rich $m_{\text{ej}}^{\text{dyn}}$ is distributed & \cite{2019MNRAS.489.5037B, shah2024predictions}\\
$A_v$ & Extinction in KN host galaxies & \cite{2009ApJS..185...32K}\\
$\mathcal{R}$\footnote{$\mathcal{R}$ is not specific to the event and is fixed for a given trial} & BNS merger rate & LVK Guide\footnote{\url{https://emfollow.docs.ligo.org/userguide/capabilities.html}  \label{url:user-guide}}\\
\bottomrule
\end{tabular}

\bigskip

\textbf{Observing run dependent parameters:}
\bigskip

\begin{tabular}{lll}
\toprule
\textbf{Input(s)} & \textbf{Description} &  \textbf{Reference(s)}\\

\midrule

\verb|Detector Uptime| & Fraction of time for which each LVK detector is operational and observing & LVK Guide$^{\ref{url:user-guide}}$, \cite{shah2024predictions}\\
\verb|Detector PSD's| & Estimated distribution of noise over frequency range of the detector & LVK Guide$^{\ref{url:user-guide}}$\\

\bottomrule
\end{tabular}

\caption{Observing run independent (top) and dependent (bottom) parameters used for simulating the time to first discoverable KN for LVK O4 and O5 runs.}
\label{table:inputs}
\end{table*}

\section{Results}\label{section:results}

\begin{figure}
    \centering
    \includegraphics[width=\linewidth]{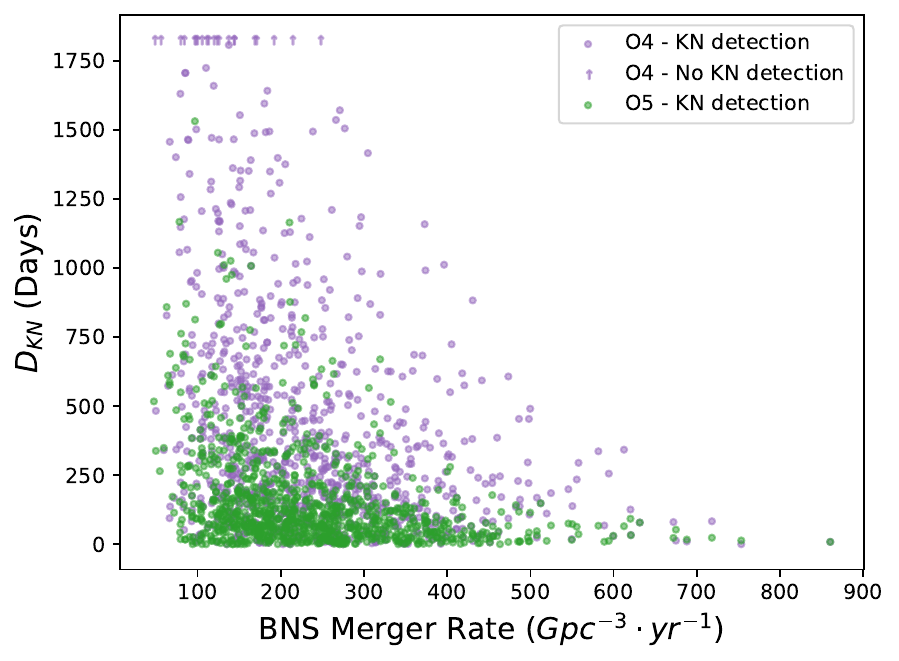}
    \caption{Scatter plot showing the day to first discoverable KN ($D_{\text{KN}}$) as a function of the BNS merger rate for LVK O4 and O5.}
    \label{fig:bns_rate_vs_days}
\end{figure}

\begin{figure}
    \centering
    \includegraphics[width=\linewidth]{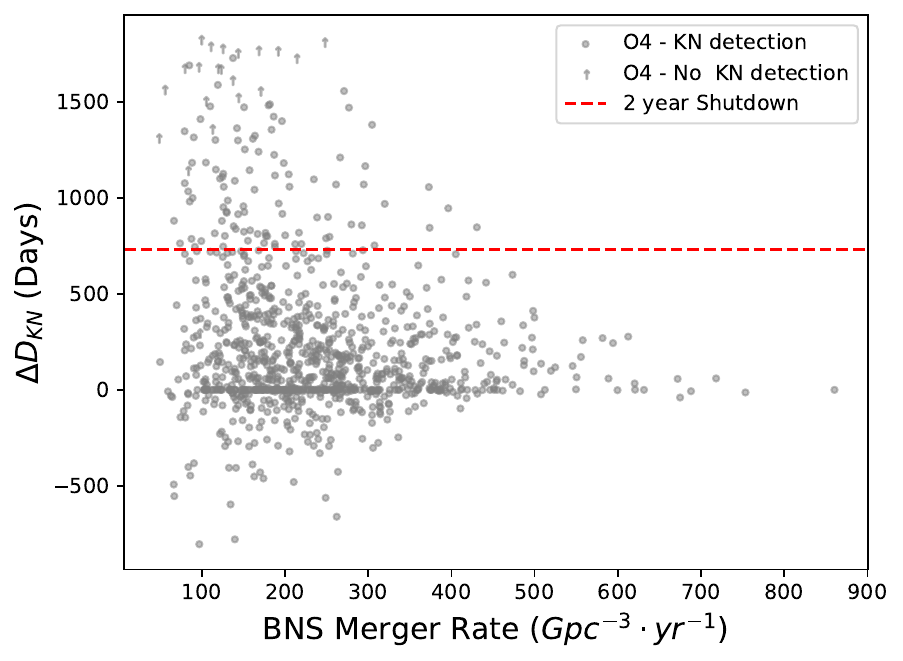}
    \caption{Scatter plot showing the difference in days to first discoverable KN between LVK O4 and O5 ($\Delta D_{\text{KN}}$) as a function of the BNS merger rate.}
    \label{fig:bns_rate_vs_delta_days_to_KN}
\end{figure}

\begin{figure}
    \centering
    \includegraphics[width=\linewidth]{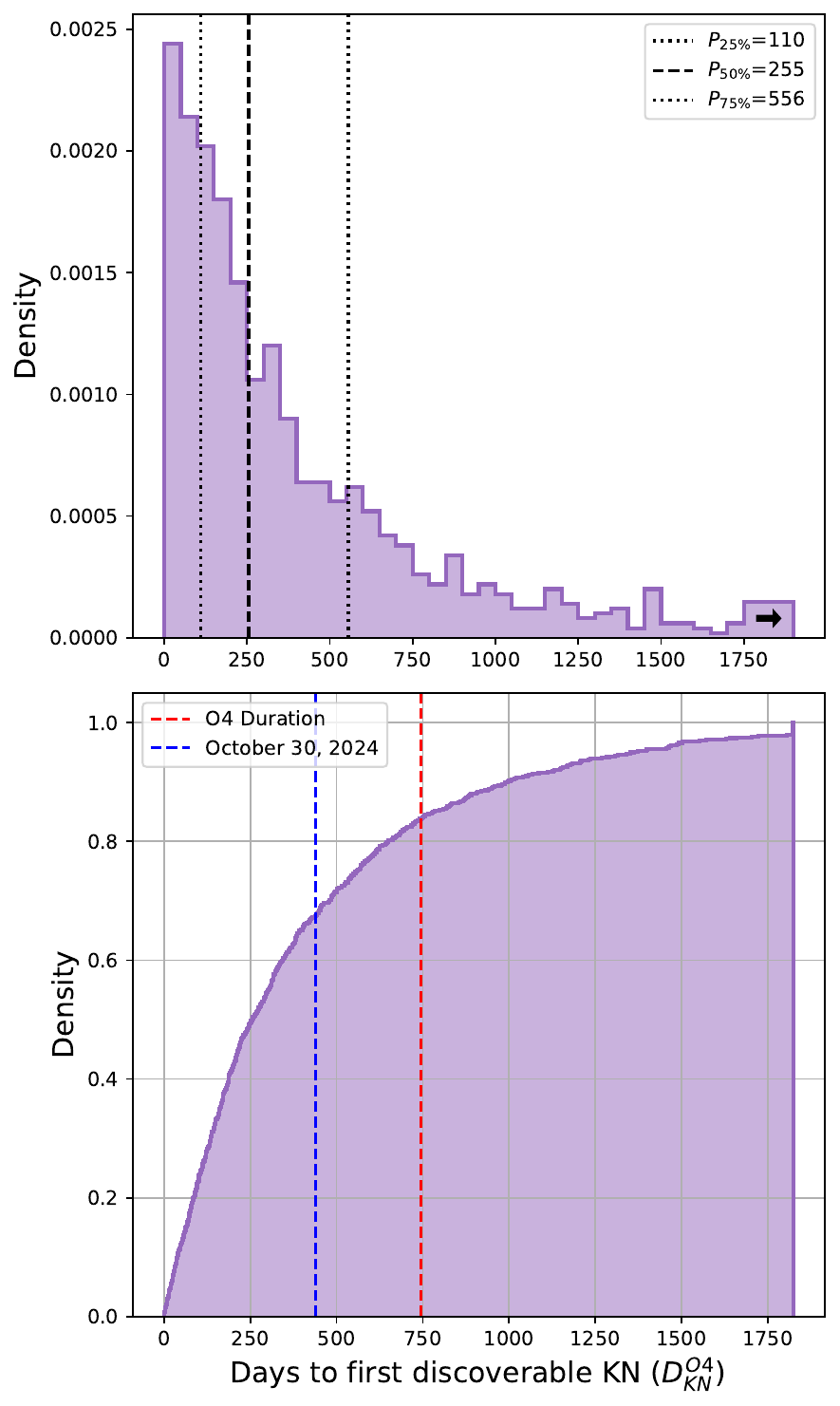}
    \caption{Distribution (top) and cumulative distribution (bottom) for the days to first discoverable KN during LVK O4 ($D_{\text{KN}}^{\text{O4}}$). The vertical blue line in the bottom panel indicates the current duration of O4 (as of October 30, 2024).  Given our priors, 68\% of our simulations had a detectable KN within the current O4 duration.}
    \label{fig:o4}
\end{figure}

\begin{figure}
    \centering
    \includegraphics[width=\linewidth]{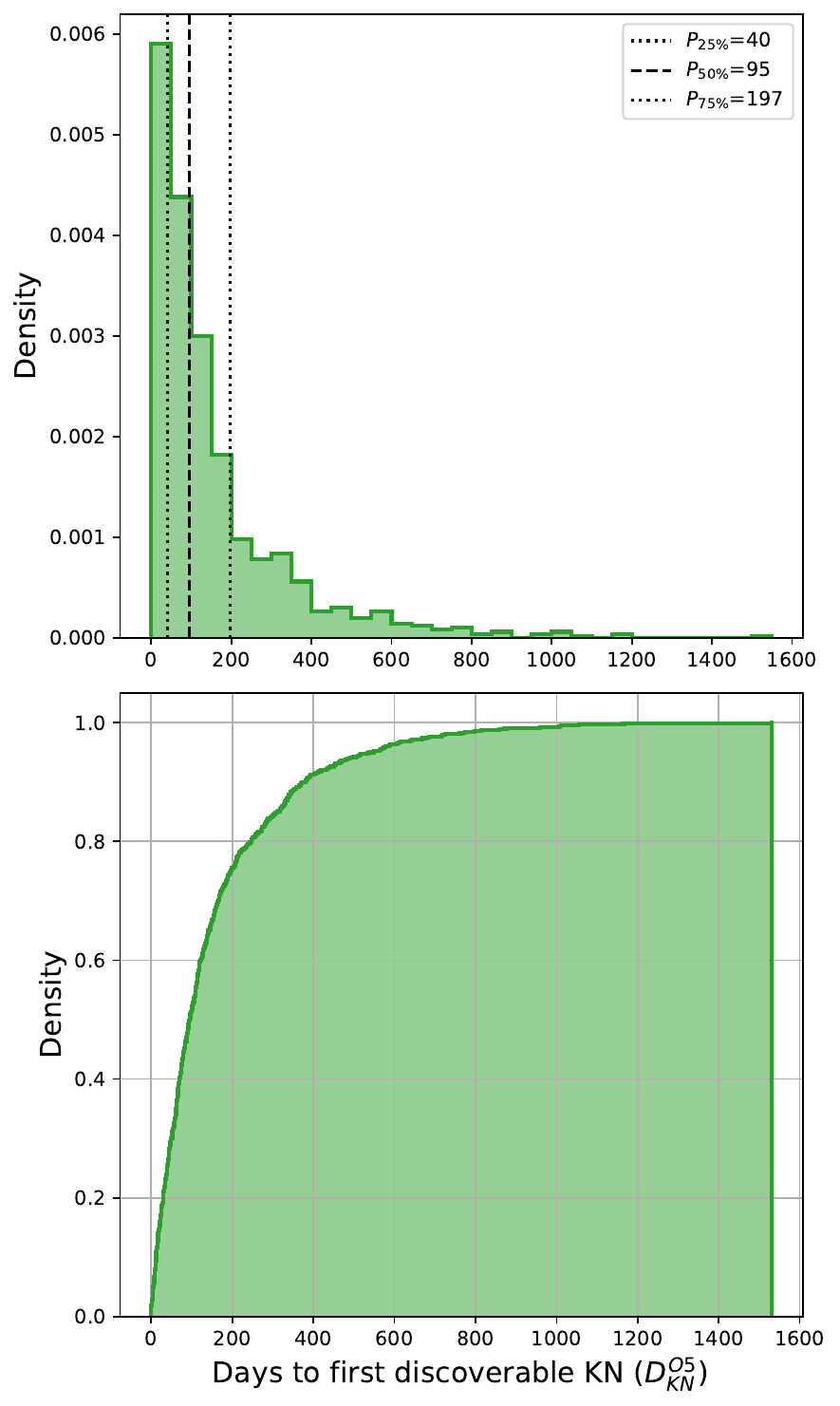}
    \caption{Distribution (top) and Cumulative Distribution (bottom) for the days to first discoverable KN during LVK O5 ($D_{\text{KN}}^{\text{O5}}$).}
    \label{fig:o5}
\end{figure}

\begin{figure}
    \centering
    \includegraphics[width=\linewidth]{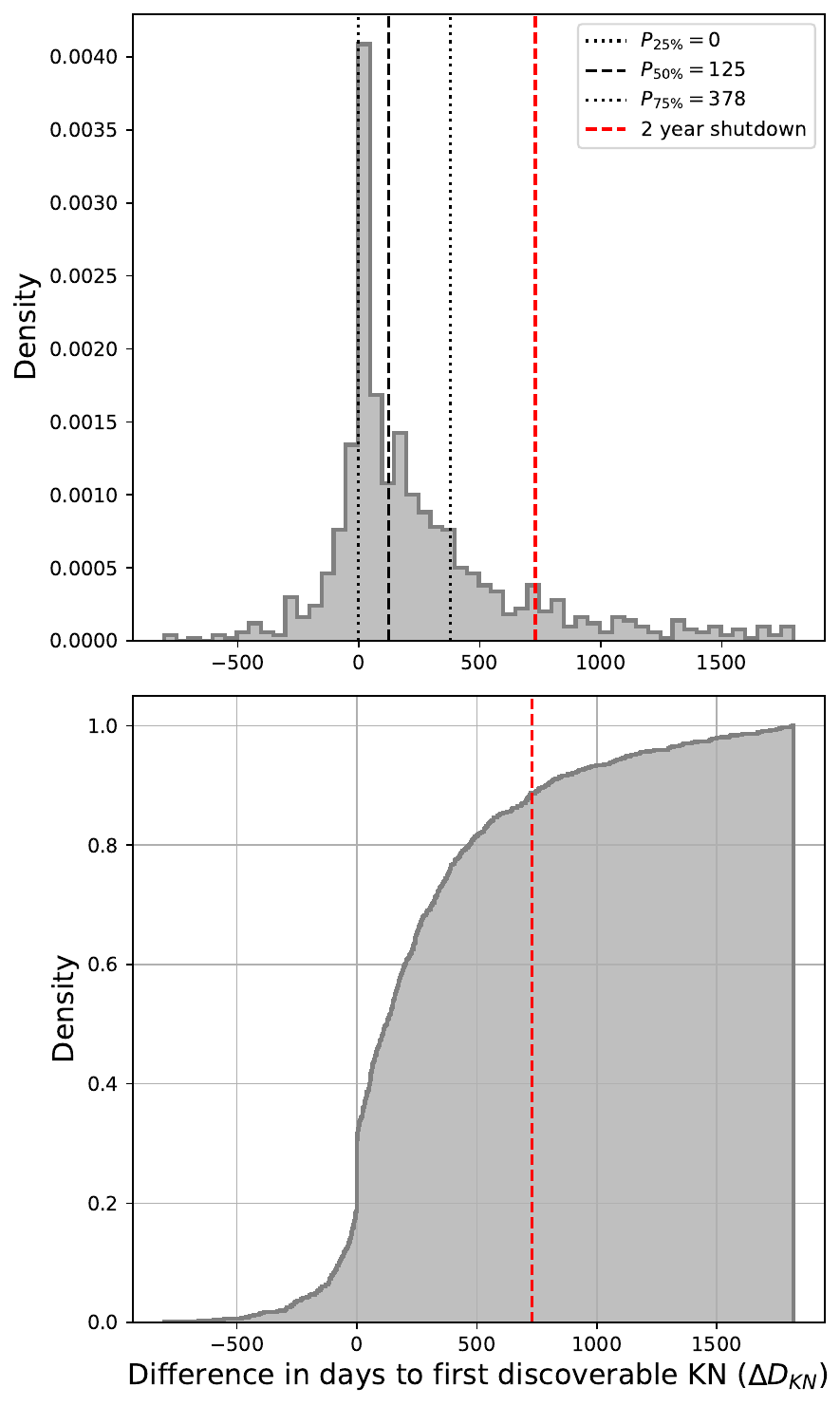}
    \caption{Distribution (top) and Cumulative Distribution (bottom) for the delta days to first discoverable KN between LVK O4 and LVK O5 ($\Delta D_{\text{KN}}$). $\Delta D_{\text{KN}} = 0$ is indicative of trials in which the first KN found during the O4 and O5 runs was produced by the same BNS system.}
    \label{fig:detla_D}
\end{figure}

\begin{figure}
    \centering
    \includegraphics[width=\linewidth]{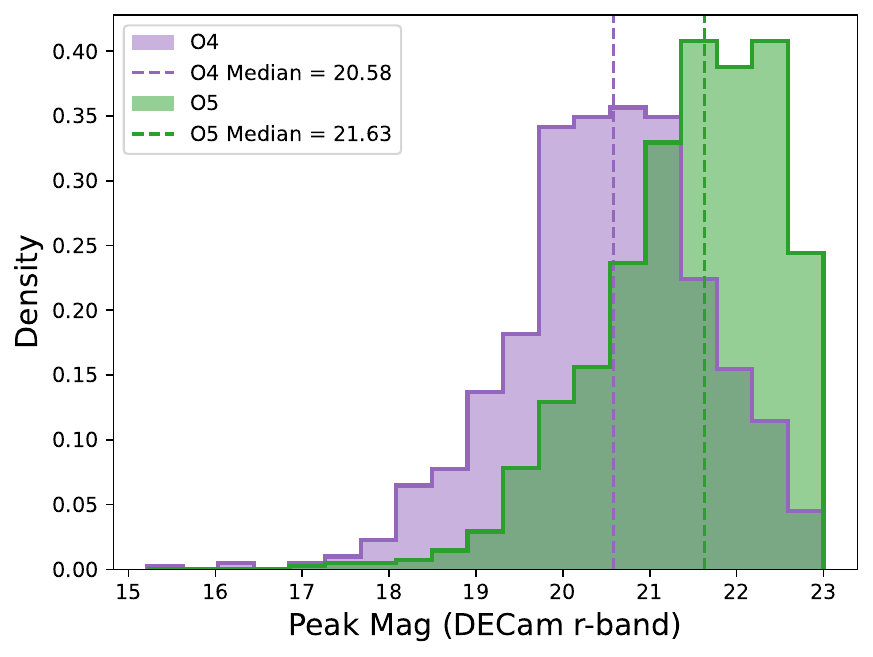}
    \caption{Distribution of the peak magnitudes for the first KN found in the LVK O4 and LVK O5 run simulations.}
    \label{fig:peak_mag}
\end{figure}

\begin{figure}
    \centering
    \includegraphics[width=\linewidth]{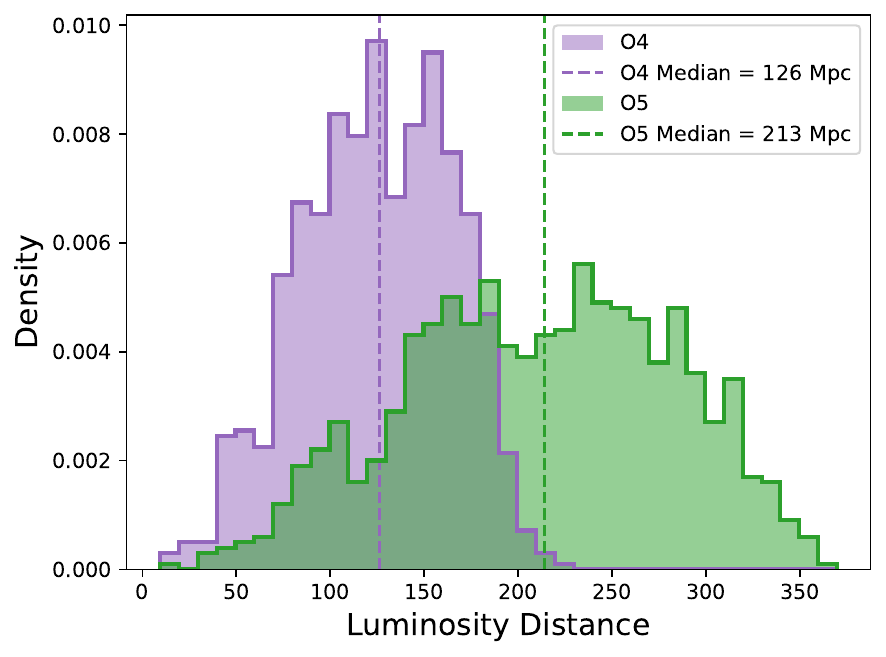}
    \caption{Distribution of the luminosity distances for the first KN found in the LVK O4 and LVK O5 run simulations.}
    \label{fig:kn_distances}
\end{figure}

Using the framework described in Sections~\ref{section:method} and \ref{section:set-up}, we performed a number of simulations.  Drawing from a set of appropriate population parameters for each trial, we produced a set of BNS merger events over a period of 5 years.  We then applied the O4 and O5 parameters to each system in a trial i.e., both observing run configurations are applied to the same set of coalescing binaries.  This allows for a direct comparison of outcomes given a specific trial.  The goal of these simulations is to measure the time from the start of an observing run until its first EM-detected KN, $D_{\text{KN}}$, and the net time difference from observing run start to first discovered KN, $\Delta D_{\text{KN}}$, with
\begin{equation}
    \Delta D_{\text{KN}} = D_{\text{KN}}^{\text{O4}} - D_{\text{KN}}^{\text{O5}}.
\end{equation}

Since both observing run configurations are applied to the same set of events in a given trial, sometimes the first detected KN is identical for both runs, resulting in a time difference of zero.  Occasionally, because of a detector being offline, a the simulation with the O4 configuration will detect a KN before the simulation with the O5 configuration for the same trial, resulting in a negative $\Delta D_{\text{KN}}$.  However because of the added sensitivity, the O5 runs typically discover a KN with a shorter delay, relative to the start of the run, than O4 runs --- as expected.

Here, we report the median and the middle $50\%$ interval for all our results. 

For LVK O4, we estimate that it would take $D_{\text{KN}}^{\text{O4}} = 255^{+301}_{-145}$ days from the beginning of the observing run to encounter the first discoverable KN. We note that in a small fraction ($\lesssim 3\%$) of LVK O4 trials, there were no discoverable KN over the 5 year period. Since we are only reporting the middle $50\%$ interval, these outliers do not affect the numbers above. For the purposes of computing $\Delta D_{\text{KN}}$, we imputed  $D_{\text{KN}}^{\text{O4}} = 1825$ days (or 5 years) for these trials. Figure \ref{fig:o4} shows the distribution and the cumulative distribution for  $D_{\text{KN}}^{\text{O4}}$.

For LVK O5, we estimate that it would take $D_{\text{KN}}^{\text{O5}} = 95^{+102}_{-55}$ days from the beginning of the observing run to encounter the first discoverable KN. Figure \ref{fig:o5} shows the distribution and the cumulative distribution for  $D_{\text{KN}}^{\text{O5}}$. It is crucial to note that these estimates use targeted LVK O5 PSDs and duty cycles. Thus, $D_{\text{KN}}^{\text{O5}}$ will increase if these sensitivity targets are not met.

We also estimate the time difference in days to first discoverable KN between LVK O4 and O5, when the the same set of run independent variables were used for both, to be $\Delta D_{\text{KN}} = 125_{-125}^{+253}$ days.  Figure \ref{fig:detla_D} shows the distribution and the cumulative distribution for $\Delta D_{\text{KN}}$.

All simulation results strongly depend on the merger rate.  Unsurprisingly, a higher rate results in a shorter time to the first KN discovery.  Although we are drawing from the current best estimate of the rate using data through O3 \citep{GWTC-3}, the lack of any BNS merger thus far in O4 suggests that the highest rates sampled here are ruled out.  We therefore present $D_{\text{KN}}^{\text{O4}}$, $D_{\text{KN}}^{\text{O5}}$ (both in Figure~\ref{fig:bns_rate_vs_days}), and $\Delta D_{\text{KN}}$ (Figure~\ref{fig:bns_rate_vs_delta_days_to_KN}) as a function of BNS merger rate.  As can be seen in those figures, there are few trials at particularly large rates (5\% at $\mathcal{R} > 450$~Gpc$^{-3}$~year$^{-1}$).  At low merger rates, the distribution of $\Delta D_{\text{KN}}$ changes some, but the qualitative results do not significantly depend on the merger rate.

The simulation outputs can be used to answer specific questions about the best approaches for KN discovery.  In particular given all the assumptions built into our simulations, we find that there is a $\sim$88\% chance that a KN will be discovered sooner by extending O4 rather than shutting down for two years and then starting O5.

\section{Conclusions}\label{section:conclusions}

Our simulations indicate that 88\% of the time, a KN will be discovered sooner if O4 is simply extended rather than shutting down for two years before starting O5.  The simulations rely on a number of assumptions including the O5 detector sensitivity and the ability to discover an EM counterpart.  Nevertheless given all data available, the current LVK plan for the O4 and O5 schedule does not appear to be optimal for discovering a KN as quickly as possible.

We can further quantify how each strategy performs.  First, the break even duration for a shutdown to upgrade the GW detectors is $125_{-125}^{+253}$~days; i.e., a shutdown shorter than $\sim 4$ months will result, on average, in a faster time to discover a KN in O5 than continuing to run at O4 sensitivity.  

Second, continuing O4 will, with 74, 88, 94, and 97\% likelihood, result in a faster EM counterpart discovery for shutdowns of 1, 2, 3, and 4 years, respectively.  Considering past LVK shutdowns have been significantly longer than expected, understanding both the expectation and contingencies is useful for decision making.

O4 started on May 24, 2023, and as of October 30, 2024, has been running for $\sim$480~days (excluding the planned 2 month break between O4a and O4b).  Virgo did not participate until March 21, 2024 and Hanford was offline for $\sim$6 weeks, so there have been only $\sim$440~days where both LIGO detectors were online.  Our simulations predict a 68\% chance of detecting a KN within that amount of observing time at O4 sensitivity.  The lack of a detection should result in a lower merger rate estimate, but barring a recalculation of the rate, we can examine the simulations where there is no detection for the initial 440~day period within an O4 trial.  Of these trials, 35\% have $\Delta D_{\text{KN}} > 2$~years, indicating that even with a lower merger rate consistent with a current non-detection, there is a 65\% likelihood that continuing O4 would resulting in a KN detection faster than shutting down for 2 years.

We have also examined the properties of the first KNe discovered in O4 and O5, with brightness and distance distributions displayed in Figures~\ref{fig:peak_mag} and \ref{fig:kn_distances}, respectively.  Since the {\it first} KN discovered in an observing run does not have particularly special properties, the distributions shown are representative of the overall populations expected to be discovered for each observing run.  As expected, those KNe discovered in O4 will be brighter (median peak $r$-band magnitude of 20.6 and 21.6, respectively) and closer (median distance of 126 and 213~Mpc, respectively) than O5.  Although we already include a brightness limit for discovery, we do not account for the full effect of brightness on detectability.  Nevertheless, brighter KNe will likely be easier to discover, be discovered earlier, and will have more extensive EM observations for a longer period.  Similarly, KNe at smaller distances, where galaxy catalogs are largely complete \citep{2022MNRAS.514.1403D}, will be easier to discover.  Larger volumes will also likely have more contaminating transients that will require vetting.  Because of these factors, the additional KNe in the O5 population will likely be harder to discover and will result in less follow-up data.

\section{Recommendations}\label{section:recommendations}

The $>$7~year delay between the first and second EM-counterpart to a GW source is an urgent issue for the field. Shutting the GW detectors down in 2025 without a new KN discovery would likely ensure a $>$10~year gap between the first and second EM counterparts.  An extended shutdown period would lead to an even longer delay, risking the postponement of critical scientific progress envisioned in the Astronomy \& Astrophysics 2020 Decadal Report to beyond 2030.

This delay also significantly affects researchers at all career stages.  The current gap between discoveries already exceeds the duration of a typical PhD program.  For instance, a student who started graduate school in 2017, just after the discovery of GW170817, likely completed their PhD (or left the field) without direct involvement in any EM counterpart discovery.  Students starting graduate studies in 2024 may have to wait until their fourth year to participate in the discovery, observations, and analysis of a KN.

In addition, various national agencies agencies may reconsider support for multi-messenger astronomy if new discoveries are separated by a decade.  Evidence of this trend is already visible as observatories have started limiting allocations in this area; notably, no {\it James Webb Space Telescope} Cycle 3 programs were approved for EM follow-up observations of LVK discoveries.  The field's viability depends on retaining personnel and knowledge, both of which are at risk if there are prolonged intervals between counterpart discoveries.

Considering the exceptional scientific opportunities a second EM counterpart would present and the significant risks associated with long discovery delays, we strongly advocate for community efforts to prioritize and mobilize resources for the fastest possible EM counterpart discovery. 

On the EM side, this requires sustained telescope allocations and enhanced coordination among different teams. On the GW side, extending O4 likely represents the most efficient path to a new counterpart discovery. 

The urgency of EM discovery is just one among many factors in making such a consequential decision. The quantitative findings of this study must be weighed against the other impacts of extended O4 operations, including research and students in other areas.  Major upgrades such as those between O4 and O5 are often subject to delays, and given the risk of a longer-than-expected delay, detecting a second EM counterpart before the shutdown presents a clear motivation for extending O4. 

\section{Acknowledgments}
\label{section:acknowledgments}

This work originated because of conversations at the 3rd TDAMM Workshop co-chaired by E.\ Burns and C.\ Fryer.  We thank D.A.\ Howell for suggesting this work be formalized as a journal article.

V.S.\ acknowledges the support of the LSST Corporation’s 2021 Enabling Science award, the 2024 Stanley Wyatt Award, and the 2024 Preble Scholarship for making this work possible. This work made use of the Illinois Campus Cluster, a computing resource that is operated by the Illinois Campus Cluster Program (ICCP) in conjunction with the National Center for Supercomputing Applications (NCSA) and which is supported by funds from the University of Illinois at Urbana-Champaign. This work was partially supported by the Center for AstroPhysical Surveys (CAPS) at the National Center for Supercomputing Applications (NCSA), University of Illinois Urbana-Champaign. 

The UCSC team is supported in part by NSF grant AST--2307710 and by a fellowship from the David and Lucile Packard Foundation to R.J.F.

GN gratefully acknowledges NSF support from AST-2206195, and a CAREER grant AST-2239364, supported in-part by funding from Charles Simonyi, AST-2432428 and OAC-2311355, DOE support through the Department of Physics at the University of Illinois, Urbana-Champaign (13771275), and support from the HST Guest Observer Program through HST-GO-16764. and HST-GO-17128 (PI: R. Foley).

\bibliography{main}{}
\bibliographystyle{aasjournal}

\end{document}